# Vortex flux and Berry phase in a Bose-Einstein condensate confined in a toroidal trap

Aranya B Bhattacherjee*

INFM, Dipartimento di Fisica E.Fermi, Universita di Pisa, Via Buonarroti 2, I-56127, Pisa, Italy

E-Mail: bhattach@df.unipi.it

**Abstract:** We study system of large number of singly quantized vortices in a BEC confined in a rotating toroidal geometry. Analogous to the Meissner effect in superconductors, we show that the external rotational field can be tuned to cancel the Magnus field, resulting in a zero vortex flux. We also show that the Berry's phase for this system is directly related to the vortex flux.

*Permanent Institute and address for correspondence: Department of Physics, Atma Ram Sanatan Dharma College, University of Delhi (South campus), Dhaula Kuan, New Delhi-110021, India.

**Introduction**

Vortices in a superfluid formed due to rapid rotation tend to organize into a triangular lattice (also known as Abrikosov lattice) as a result of mutual repulsion. Such a lattice also has an associated rigidity. In the recent years, it has become possible to achieve such a vortex lattice state in a rapidly rotating Bose-Einstein condensate [1]. These new developments in experiments have opened a new direction for the theoretical study of quantum vortex matter, such as fractional quantum Hall like and melted vortex lattice states [2-7]. This new state of quantum vortex matter occurs only at very rapid rotations, where the system enters the " mean field" quantum Hall regime, in which the condensation is only in lowest Landau level orbits. Eventually the vortex lattice melts, and the system enters a strongly correlated regime. If a vortex moves with respect to a liquid, classical or quantum, there is a force on the vortex normal to the relative vortex velocity with respect to the liquid. This is the Magnus force which determines the mutual friction in superfluids [8-10] and the Hall effect in superconductors [11].

Ao and Thouless [12] have pointed out that the Magnus force is connected with the Berry phase [13], which is the phase variation of the quantum-mechanical wave function of a quantum liquid generated by the adiabatic transport of the vortex round a close loop. From the Berry phase analysis Ao and Thouless [12,14] concluded that the effective Magnus force is proportional to the superfluid density. In this work, we study the dynamics of a large number of singly quantized vortices in a rotating Bose-Einstein condensate confined in a toroidal geometry. In particular, we study the effective Hamiltonian for the vortex degrees of freedom, motivated by an analogy [4] between the Magnus force acting on a vortex moving in a two-dimensional neutral superfluid and the Lorentz force acting on a charged particle in a magnetic field.

**Magnus force and vortex flux**

We consider a large number of vortices in a rotating gas of bosons confined in a toroidal trap at temperatures well below the Kosterlitz-Thouless transition temperature given by $k_B T_C^{KT} = \hbar^2 \rho_s / 4m^2$, where $\rho_s$ is the mass density and $m$ is the mass of the single bosonic atom. For theoretical ease, we consider a narrow torus whose cross-sectional area $A$ is small compared with the square of its radius $R$. Consequently, we can regard the system as a ring of radius $R$. Such a kind of toroidal traps has been extensively studied by many workers [15,16]. The BEC is described by a repulsive short range interaction $V(\vec{r}) = g_{2D} \delta^2(\vec{r})$, with the 2D interaction parameter $g_{2D} = \sqrt{8\pi} \hbar^2 a_s / m a_z$, where $a_s$ is the s-wave scattering length. The superfluid forms a triangular lattice of quantized vortices (carrying the angular momentum of the system) rotating as a solid body at angular velocity $\Omega$. We assume that at low temperatures, the scattering of thermal excitations by vortices does not affect the vortex dynamics. The situation considered then corresponds to considering a vortex as a point particle moving under the influence of the Magnus force.

The vortex motion is governed by a Hamiltonian corresponding to one of point particles, with a charge equal to the quantum of circulation $\kappa = h/m$, interacting with electromagnetic fields. For a large collection of vortices $N_V \gg 1$ in a frame rotating with angular velocity $\Omega$, the vortex Hamiltonian reads:

$$H_V = \sum_{i=1}^{N_V} \frac{(\vec{P}_i - \kappa\vec{A}_i)^2}{2m_v} - \Omega(\vec{X}_i \times \vec{P}_i)_z - \frac{\rho_S \kappa^2}{2\pi} \sum_{i<j}^{N_V} \ln\left|\frac{\vec{X}_i - \vec{X}_j}{\xi}\right| \tag{1}$$

The first term in Eqn.(1) is the kinetic energy of the vortices, the second term is a result of centrifugal and coriolis forces on vortices and the last term is the repulsive logarithmic Coulomb interaction of point vortices. $\xi$ is the coherence length and is $\xi = \sqrt{\hbar^2/2g_{2D}\rho_S}$. The effective vortex mass $m_v = \pi\rho_S\xi^2$ [4]. The pseudo vector potential due to the Magnus force is $A_i^a = \frac{1}{2}\rho_S \varepsilon^{ab} X_b^i$ ($a,b = x,y$) [4]. We can rewrite eqn.(1) as

$$H_V = \sum_{i=1}^{N_V} \frac{(\vec{P}_i - \kappa\vec{A}_i - \kappa\vec{a}_i)^2}{2m_v} + \frac{\hbar^2}{2m_v R^2} - \frac{1}{2}\Omega R^2(m_v\Omega - \kappa\rho_s) - \frac{\rho_S \kappa^2}{2\pi} \sum_{i<j}^{N_V} \ln\left|\frac{\vec{X}_i - \vec{X}_j}{\xi}\right| \tag{2}$$

Where, $\vec{a}_i = \frac{m_v}{\kappa}(\vec{\Omega} \times \vec{X}_i)$ and in a torus $x_i^2 + y_i^2 = R^2$.

We define the average value of the vortex flux as

$$\vec{\bar{J}}_{vor} = \vec{\nabla} \times \kappa(\vec{\bar{A}} + \vec{\bar{a}}) \tag{3}$$

$(\vec{\bar{A}} + \vec{\bar{a}}) = \vec{A}_{eff}$ is the effective vector potential. Substituting the expressions for $\vec{A}$ and $\vec{a}$, we find

$$\vec{\bar{J}}_{vor} = (2\Omega m_v - \kappa\bar{\rho}_s)\hat{z} \tag{4}$$

$\bar{\rho}_s$ is the average value of $\rho_s$. The average value of the flux vanishes for a critical $\Omega = \Omega_C = \kappa\rho_s/2m_v = \hbar/m_v R^2$. It is to be noted that all the vortices will be subjected to the same Coriolis force only in a tight torus.

We notice that the vortices interact with the sum of an external field $\vec{a}_\mu$ (related to the rotation of the trap and $\mu = x,y$) and the Magnus field $\vec{A}_\mu$. $\vec{A}_\mu$ is an internal degree of freedom that is correlated to the fluctuations of the particle density $\rho_s$. We represent it as the sum of the average value $\vec{\bar{A}}_\mu$ and fluctuations $\vec{\delta A}_\mu$ around it, $\vec{\bar{A}}_\mu + \vec{\delta A}_\mu$. We write for $\vec{a}_\mu$ the sum of $\vec{\bar{a}}_\mu$ corresponding to a constant rotation $\Omega$ along the $z$ direction and an infinitesimal test field $\vec{\delta \tilde{a}}_\mu$ that is introduced to study the linear response of the system, $\vec{\bar{a}}_\mu + \vec{\delta \tilde{a}}_\mu$. Analogous to the Meissner effect in superconductors, the external rotation field $\vec{a}_\mu$ can be tuned to cancel the Magnus field $\vec{A}_\mu$, resulting in a zero vortex flux.

**Effective vector potential and the Berry phase**

Berry [13] showed that as the Hamiltonian of a system is varied adiabatically through a complete cycle in its parameters, a phase (Berry's phase), in addition to the usual dynamical phase, develops for the state vector. The Berry's phase for the rotating vortex lattice is formed to be:

$$\gamma = \frac{1}{\hbar} \oint \vec{A}_{eff} \cdot d\vec{X} = -\frac{1}{\hbar} \iint (\vec{\nabla}_L \times \vec{A}_{eff}) d\vec{S} = n\pi \left\{ 1 - \frac{\Omega}{\Omega_C} \right\} \qquad (5)$$

Where $\Omega_C = \hbar / m_v R^2$, $\vec{\nabla}_L$ is the Laplacian w.r.t the vector position vector $\vec{X}_L$. Eqn.(5) reveals that the Berry's phase vanishes as $\Omega \to \Omega_C$.

**Acknowledgements:** I acknowledge support by the Abdus Salam International Centre for Theoretical Physics, Trieste, Italy under the ICTP-TRIL fellowship scheme. My thanks to Prof. E. Arimondo, for providing me the facilities for this work at INFM, Pisa.

**Conclusions**

In summary, we have shown that for a system comprising of a large number of vortices in a rotating BEC confined in a torus, there is a competition between the internal degree of freedom (the Magnus field) and the external degree of freedom (the rotational field). The vortex flux vanishes (i.e the vortex rotation around the torus stops) when the two fields balances. This can be accomplished by tuning the external rotational frequency of the torus. This effect is analogous to the Meissner effect in superconductors. The berry's phase is found to be proportional to the vortex flux.